\documentclass[a4paper]{jpconf}

\usepackage{nicefrac}
\usepackage{pst-all,graphicx}

\begin{document}

\title{Calculations of direct photon emission in Heavy Ion Collisions at
$\sqrt{s_{\rm NN}} = 200$~GeV}

%\ShortTitle{Direct Photons In Heavy Ion Collisions}

\author{B~B\"auchle~and~M~Bleicher}
\address{Frankfurt Institute for Advanced Studies, Ruth-Moufang-Stra\ss{}e
1, 60438 Frankfurt am Main,
        Germany}
\ead{baeuchle@th.physik.uni-frankfurt.de}
\ead{bleicher@th.physik.uni-frankfurt.de}

\begin{abstract}
Direct photon emission in heavy-ion collisions is calculated within a
relativistic micro+macro hybrid model and compared to the microscopic
transport model UrQMD.  In the hybrid approach, the high-density part of the
collision is calculated by an ideal 3+1-dimensional hydrodynamic
calculation, while the early (pre-equilibrium-) and late (rescattering-)
phase are calculated with the transport model. We study both models with
$Au+Au$-collisions at $\sqrt{s_{\rm NN}} = 200$~GeV and
compare the results to experimental data published by the PHENIX
collaboration.
\end{abstract}

\section{Introduction}

Creating and studying high-density and -temperature nuclear matter is the
major goal of heavy-ion experiments. A state of quasi-free partonic degrees
of freedom, the Quark-Gluon-Plasma (QGP)~\cite{Harris:1996zx,Bass:1998vz}
may be formed, if the energy density reached in the reaction is high enough.
Strong jet quenching, large elliptic flow and other observations made at the
Relativistic Heavy Ion Collider (BNL-RHIC) suggest the successful creation
of a strongly coupled QGP (sQGP) at these
energies~\cite{Adams:2005dq,Back:2004je,Arsene:2004fa,Adcox:2004mh}, and
possible evidence for the creation of this new state of matter has also been
put forward by collaborations at the Super Proton Synchrotron (CERN-SPS), as
for instance the step in the mean transverse mass excitation function of
protons, kaons and pions and the enhanced $K^+/\pi^+$-ratio \cite{:2007fe}.

Electromagnetic probes provide a unique insight into the early stages of
heavy-ion collisions, since they have the advantage of negligible
rescattering cross-sections. Therefore, they leave the production region
without rescattering and carry the information from this point to the
detector. Besides single- and dileptons, direct photon emission can
therefore be used to study the early hot and dense, possibly partonic,
stages of the reaction.

Unfortunately, most photons measured in heavy-ion collisions come from
hadronic decays. The experimental challenge of obtaining spectra of only
direct photons has been gone through by several experiments; WA98
(CERN-SPS)~\cite{Aggarwal:2000ps} and PHENIX
(BNL-RHIC)~\cite{Adler:2005ig} have published explicit data points
for direct photons.

On the theory side, the elementary photon production cross-sections are
known since long, see e.g.\ Kapusta {\it et al.}~\cite{Kapusta:1991qp} and
Xiong {\it et al.}~\cite{Xiong:1992ui}.  The major problem is the difficulty
to describe the time evolution of the produced matter, for which first
principle calculations from Quantum Chromodynamics (QCD) cannot be done.
Well-developed dynamical models are therefore needed to describe the
space-time evolution of nuclear interactions. 

Among the approaches used are relativistic transport
theory~\cite{Geiger:1997pf,Bass:1998ca,Bleicher:1999xi,Ehehalt:1995is,Molnar:2004yh,Xu:2004mz,Lin:2004en,Burau:2004ev,Bass:2007hy}
and relativisitc fluid- or
hydrodynamics~\cite{Cleymans:1985wp,McLerran:1986nc,PHRVA.D34.794,Kataja:1988iq,Srivastava:1991ju,Srivastava:1991nc,Srivastava:1991dm,Srivastava:1992gh,Cleymans:1992zc,Rischke:1995mt,Hirano:2001eu,Huovinen:2001wx,Huovinen:2002im,Kolb:2003dz,Nonaka:2006yn,Frodermann:2007ab}.
For both models, approximations have to be made, and in both models, the
restrictions imposed by these approximations can be loosened. For transport
theory, the necessary approximations include the restriction of scattering
processes to two incoming particles, which limits the applicability to low
particle densities. For hydrodynamics, on the other hand, matter has to be
in local thermal equilibrium (for ideal, non-viscous hydrodynamic
calculations) or at least close to it (for viscous
calculations)~\cite{Dusling:2007gi,Baier:2006um,Song:2008si}.

From this, it is clear where the advantages for both models
are: While in transport, non-equilibrium matter, which is present in the
beginning of the heavy-ion reaction, and dilute matter, which is present in
the late phase, can be described, hydrodynamics may be better suited to
describe the intermediate stage, which is supposed to be dense, hot and
thermalized. In addition, the transition between two phases of matter, such
as Quark Gluon Plasma (QGP) and Hadron Gas (HG) can be easily described in
hydrodynamics, while this is not (yet) possible for transport models, since
the microscopic details of this transition are not known.

\section{The Model}\label{sec:model}

UrQMD v2.3 (Ultra-relativistic Quantum Molecular Dynamics) is a microscopic
transport model~\cite{Bass:1998ca,Bleicher:1999xi,Petersen:2008kb}. It includes all
hadrons and resonances up to masses $m \approx 2.2~{\rm GeV}$ and at high
energies can excite and fragment strings. The cross-sections are either
parametrized, calculated via detailed balance or taken from the additive
quark model (AQM), if no experimental values are available.  At high parton
momentum transfers, PYTHIA~\cite{Sjostrand:2006za} is employed for pQCD
scatterings. 

UrQMD differentiates between two regimes for the excitation and
fragmentation of strings. Below a momentum transfer of $Q < 1.5$~GeV a
maximum of two longitudinal strings are excited according to the LUND
picture, at momentum transfers above $Q > 1.5$~GeV hard interactions are
modelled via PYTHIA. For detailed information on the inclusion of PYTHIA,
the reader is referred to Section~II of~\cite{Petersen:2008kb}. In the UrQMD
framework, propagation and spectral functions are calculated as in vacuum.

In the following, we compare results from this microscopic model to results
obtained with a hybrid model description~\cite{Petersen:2008dd}. Here, the
high-density part of the reaction is modelled using ideal 3+1-dimensional
fluid-dynamics.  The unequilibrated initial state and the low-density final
state are described by UrQMD. In the hydrodynamic intermediate stage we use
a Hadron Gas Equation of State (HG-EoS) which includes the same degrees of
freedom as are present in the transport phase. This allows to explore the
effects due to the change of the kinetic description.

To connect the initial transport phase with the high-density fluid phase,
the baryon-number-, energy- and momentum-densities are smoothed and put into
the hydrodynamic calculation after $t = 0.6$~fm. Temperature, chemical
potential, pressure and other macroscopic quantities are determined from the
densities by the Equation of State used in the current calculation. During
this transition, the system is forced into an equilibrated state. Particles
with high rapidity $y > 2$, however, are excluded from the hydrodynamic grid
and propagated in the cascade without interaction to the hydrodynamic
medium.

In non-central collisions, the spectators are propagated in the cascade.
After the local rest frame energy density has dropped below a threshold
value of $\epsilon_{\rm crit} \approx 5 \epsilon_0$, particles are created
on a hyper-surface from the densities by means of the Cooper-Frye formula
and propagation is continued in UrQMD. 

The transition from hydrodynamic to cascade description used in the
calculations presented here is gradual. I.e. each transverse slice (constant
$z$) is transferred to the cascade at the same time, when the condition is
met throughout that slice.  This represents a pseudo-eigentime condition.

Photon emission is calculated perturbatively in both models, hydrodynamics
and transport, because the evolution of the underlying event is not altered
by the emission of photons due to their very small emission probability. The
channels considered for photon emission may differ between the hybrid
approach and the binary scattering model.  Emission from a
Quark-Gluon-Plasma can only happen in the hydrodynamic phase, and only if
the equation of state used has partonic or chirally restored degrees of
freedom. Photons from baryonic interactions are neglected in the present
calculation.

For emission from the transport part of the model, we use the
well-established cross-sections from Kapusta {\it et
al.}~\cite{Kapusta:1991qp}, and for emission from the hydrodynamic phase, we
use the parametrizations by Turbide, Rapp and Gale and Arnold {\it et
al.}~\cite{Turbide:2003si} (the latter for QGP-emission).  For detailed
information on the emission process, the reader is referred to B\"auchle and
Bleicher~\cite{arXiv:0905.4678}.

\section{Results}

Photon emission has been calculated for minimum bias (0-92~\%) and central
(0-10~\%) $Au+Au$-collions at $\sqrt{s_{\rm NN}} = 200$~GeV with both the
pure cascade and the cascade-hydrodynamic hybrid models. We note that the
high-$p_\bot$ data obtained by PHENIX~\cite{Adler:2005ig} can be described
very well with pQCD-calculations by Gordon and
Vogelsang~\cite{Gordon:1993qc}, therefore we investigate the contributions
of our model to low-$p_\bot$ photons only.

\begin{figure}
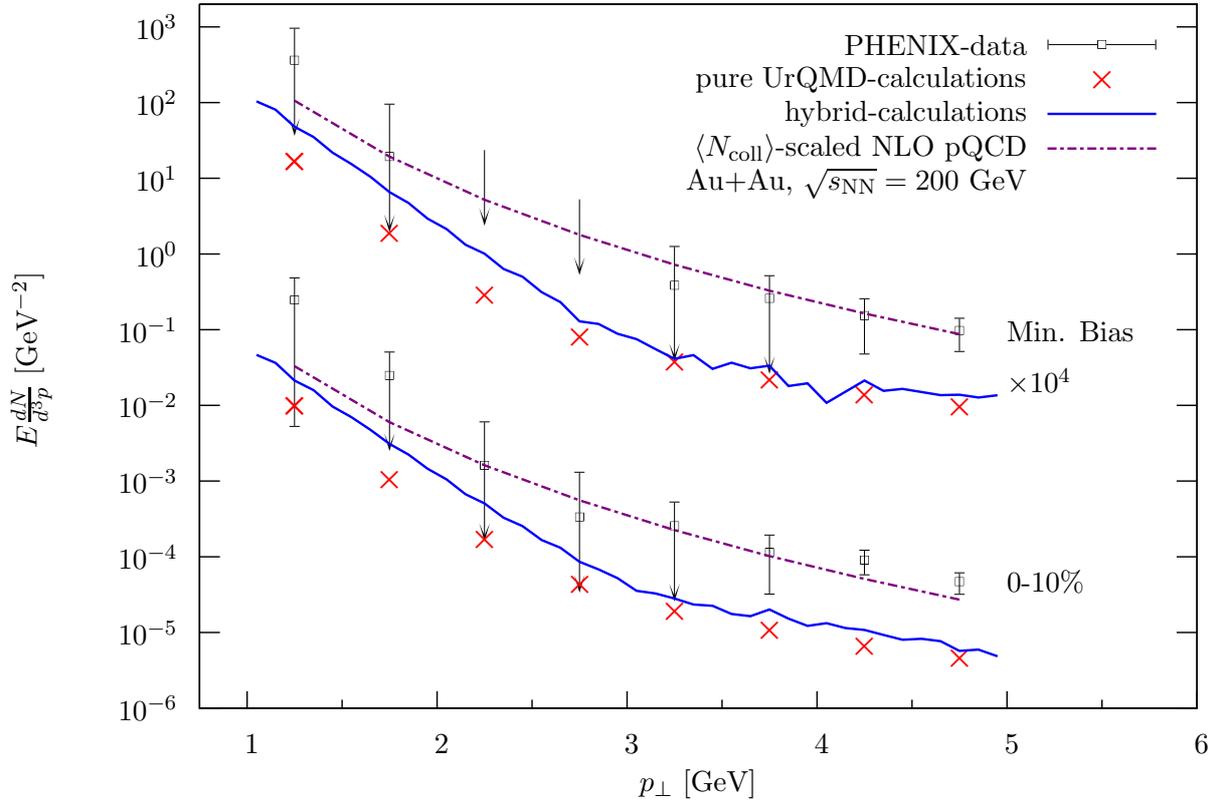

 \include{phenix_central}
 \caption{Comparison of the direct photon spectra from UrQMD cascade- (red
 crosses) and hybrid- (blue solid line) calculations to the spectra measured
 by the PHENIX collaboration (black symbols) and to the pQCD contribution
 predicted by Gordon and Vogelsang (violet dash-dotted line) for minimum
 bias and central Au+Au-collisions at $\sqrt{s_{\rm NN}} = 200$~GeV.}
 \label{fig:phenix_low}
\end{figure}

In Figure~\ref{fig:phenix_low}, we show the direct photons emitted by the
cascade- and hybrid models and compare them to the aforementioned
pQCD-calculations and data from the PHENIX-collaboration.

While our hadronic model fails to predict the amount of direct photons
emitted at intermediate transverse momentum, the disagreement is smaller at
low $p_\bot$. The spectra obtained with an intermediate hydrodynamic part
are significantly higher than those with pure cascade calculations. In any
case, the spectra are negligible compared to the pQCD contributions
predicted by Gordon and Vogelsang, therefore, an enhancement with respect to
those data cannot be explained by hadronic sources.

\begin{figure}
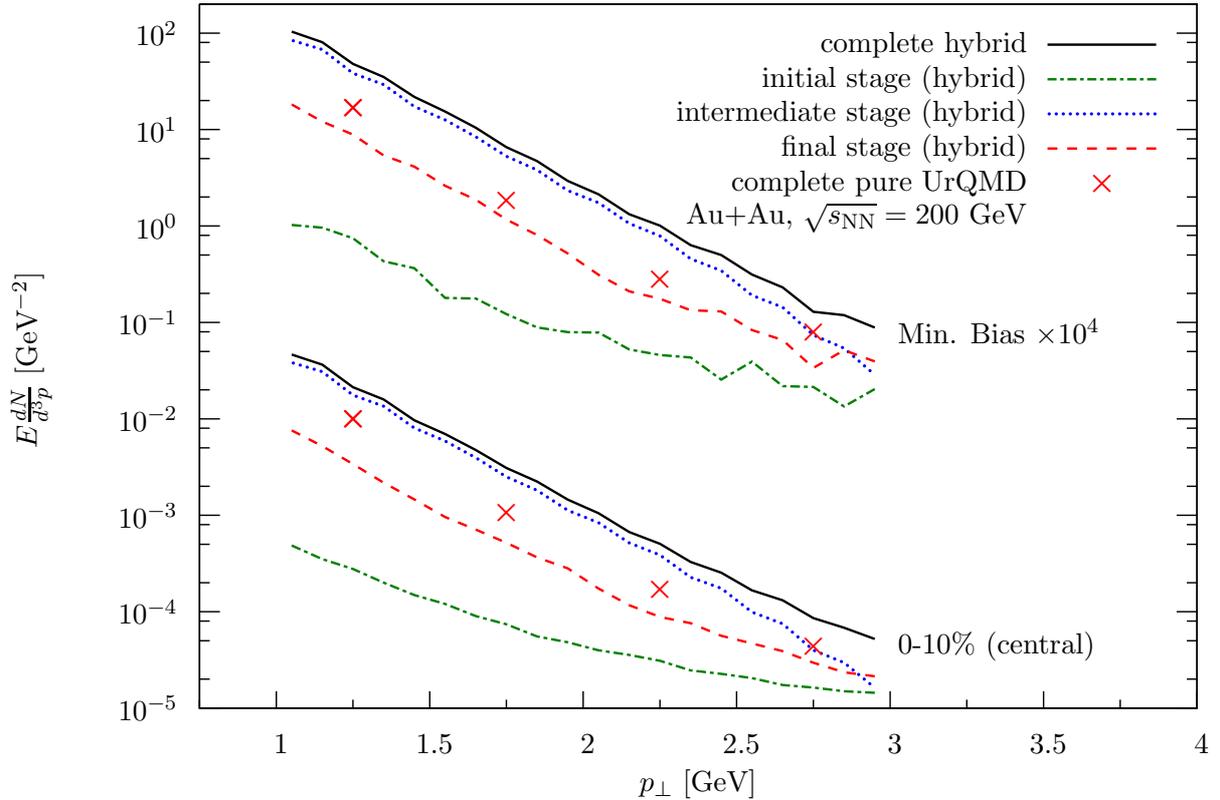

 \include{phenix_stages}
 \caption{Comparison of the contribution of the different stages (initial
 stage, green dash-dotted lines; intermediate stage, blue dotted lines,
 final stage, red dashed lines) to the overall spectrum from
 the hybrid model (black solid line) and the overall spectrum from pure
 cascade calculations (red crosses) for minimum bias and central
 Au+Au-collisions at $\sqrt{s_{\rm NN}} = 200$~GeV.}
 \label{fig:stages}
\end{figure}

In Figure~\ref{fig:stages}, we explore the contribution of the different
stages to the hybrid model spectra. The contribution of the intermediate
hydro phase turns out to be dominant at low transverse momenta $p_\bot <
3$~GeV. The contributions of the initial and final stages are lower than the
complete spectrum from the cascade calculations, so that we can assume the
intermediate stage in the cascade calculation to be on the order of the
final stage contribution in the hybrid model calculations.

The difference between the hydrodynamic and transport descriptions of the
intermediate stage has not been observed for collisions with lower energies
$E_{\rm lab} = 158$~AGeV (see~\cite{arXiv:0905.4678}) and $E_{\rm lab} =
45$~AGeV (see~\cite{arXiv:1003.5454}).

\section{Summary}

In this article, we have applied UrQMD and the UrQMD+Hydro hybrid model to
calculate photon spectra from central and minimum bias Au+Au-collions at
$\sqrt{s_{\rm NN}} = 200$~GeV. The comparison of the (hadronic) calculations
to data from the PHENIX collaboration suggest that a significant
contribution to the measured spectra comes from non-hadronic sources. The
low-$p_\bot$-excess over pQCD-predictions seen by the
PHENIX-collaboration~\cite{:2008fqa} cannot be explained by hadronic
sources; partonic sources such as a Quark-Gluon-Plasma are therefore very
likely to be responsible for this excess.

The comparison of transport and hybrid calculations show that the conclusion
drawn at lower energy, which suggested that there is no difference between
the spectra obtained with or without intermediate hydrodynamic stage, is not
valid at these high energies. The excess of the hybrid model calculations
over the transport calculations are visible in both central and minimum bias
collision samples, and the magnitude of this excess is similar in both
models. This suggests that the excess depends only the collision energy, not
on the system size.

\section{Outlook}

The results shown here suggest the need for further studies. Calculations of
direct photon spectra have to be done with the current model for other
centrality selections, and with different Equations of State such as a Bag
Model EoS and an EoS with a chirally restored phase. In order to investigate
the difference between the hybrid model and the cascade model, an energy
dependent investigation of the excess is advisable.

Different systems measured at the RHIC-facility, such as Cu+Cu-collisions
and collisions at smaller center-of-mass energy, will also be calculated.

The parameters of the model -- the conditions for switching to and from the
hydrodynamic description and the scenario for the latter transition -- will
be investigated in the future.

\ack

This work has been supported by the Frankfurt Center for Scientific
Computing (CSC), the GSI and the BMBF. The authors thank Hannah Petersen for
providing the hybrid- and Dirk Rischke for the hydrodynamic code. B.\
B\"auchle gratefully acknowledges support from the Deutsche Telekom
Stiftung, the Helmholtz Research School on Quark Matter Studies and the
Helmholtz Graduate School for Hadron and Ion Research. This work was
supported by the Hessian LOEWE initiative through the Helmholtz
International Center for FAIR.

The authors thank Elvira Santini, Pasi Huovinen and Rene Bellwied for
valuable discussions and Klaus Reygers for experimental clarifications.

\end{document}